\documentclass[journal=ancac3,manuscript=article]{achemso}

\author{Jason Hao}
\affiliation{Thomas Jefferson High School for Science and Technology, Alexandria, VA 22312, USA}
\author{Jeffrey Owrutsky}
\affiliation{Precise Systems, Inc., Patuxent River, MD 20653, USA}
\author{Daniel Ratchford}
\affiliation{U.S. Naval Research Laboratory, Washington D.C. 20375, USA}
\author{Blake Simpkins}
\affiliation{U.S. Naval Research Laboratory, Washington D.C. 20375, USA}
\author{Alexander L. Efros}
\email{*alex.l.efros.civ@us.navy.mil} 
\affiliation{U.S. Naval Research Laboratory, Washington D.C. 20375, USA}

\title{Surface Exciton Polariton}

\keywords{Surface Exciton Polariton, Maxwell's Equations, Analytical Solutions, Dispersion Relations}

\usepackage{amsmath}
\usepackage{amssymb}
\usepackage{bm}
\usepackage{graphicx}
\usepackage{color}
\usepackage{hyperref}

\definecolor{blueviolet}{rgb}{0.54,0.17,0.89}

\def\be{\begin{equation}}
\def\ee{\end{equation}}
\def\bea{\begin{eqnarray}}
\def\eea{\end{eqnarray}}

\begin{document}
\maketitle

\begin{abstract}
In this Letter, we  have developed a theory describing surface exciton polariton (SEPs) that accounts for the spatial dispersion of the dielectric constant connected with exciton momentum. Due to strong coupling between light and bulk excitons in the frequency separation, $\hbar\omega_{LT}$,  between the  longitudinal  and transverse exciton,  the SEP  is formed and behaves at partially light and partially matter.  The dispersion  of the SEP was found through a combined solution of Maxwell's and Thomas-Hopfield's equations. The analytical theory  describes  SEPs at any  bulk exciton/vacuum interface  and provides its complete dispersion  if one knows $\hbar\omega_{LT}$, the  exciton effective mass, $M$, and the  high frequency dielectric constant, $\kappa_\infty$. The presented theory is in excellent  agreement with the only numerical  modeling of this problem, which was conducted  for SEPs at a ZnO/vacuum interface. Calculations show the spatial dispersion of the dielectric constant  leads to rather small broadening  of  the photon-like quasi-particle and  suggests using SEPs for long-range coherence transfer.  
\end{abstract}

\maketitle

Surface polaritons, such as plasmon-polaritons, exciton-polaritons, and phonon polaritons are bosonic quasi-particles that are part light and part matter, arising from strong coupling of bulk material excitations and photonic modes at the matter/vacuum  surface. They have attracted attention for their potential in transferring coherence over long distances through their photonic component and conducting computations or operations with the material part of their wave function. In addition, the wavelength of these quasi-particles can be significantly smaller than the free-space wavelength, allowing one to create devices with sizes below the diffraction limit.  

In general, surface polaritons are formed in the frequency range where the real part of the dielectric constant of the material is negative: $\kappa(\omega)<0$, where $\omega$ is the wave frequency. Solving Maxwell's equations  and using standard boundary conditions, one may obtain (see SI)  the wave that propagates along the surface with wave vector $q$ and exponentially decays into both sides of the interface (i.e., the solid and vacuum). The square of the wave vector of these surface waves satisfies the following equations
\be
q^{2} = \frac{\omega^{2}}{c^{2}}\left(\frac{\kappa(\omega) }{1+\kappa(\omega)}\right)
\label{eq:1}
\ee
where $c$ is the speed of light. One can see that the surface wave exists ($q^2>0$) when $\kappa(\omega)<-1$. Equation \eqref{eq:1} also shows that the speed of the surface waves, $v_s$, is smaller than the speed of light: $c\sqrt{|\kappa(\omega)|-1]/|\kappa(\omega)|}<c$.

Equation \eqref{eq:1} can be  used to describe surface plasmon polaritons.\cite{book}. For a metal, the dielectric constant  $\kappa^m(\omega)= 1-\omega_{p}^{2}/\omega^{2}$, is negative in the frequency range $0<\omega<\omega_p$, where $\omega_p$ is the plasmon frequency. Substituting  $\kappa^m(\omega) $ in Eq. \eqref{eq:1}, one can obtain the well know dispersion of the  surface plasmon polariton:
$\omega= \omega_{p}\sqrt{(q^{2}c^{2}-1)/(2c^{2}q^{2}-1)}$

Surface exciton polaritons (SEPs) can be created at  the  surface of a semiconductor in the frequency range between  the transverse, $\omega_T$, and longitudinal, $\omega_L$, exciton  frequencies:  $\omega_T < \omega <\omega_L=\omega_T+\omega_{LT}$, where the dielectric constant is negative and $\omega_{LT}$ is the longitudinal-transverse splitting of the exciton spectra.  In contrast to the dielectric constant of a metal, the semiconductor exhibits a dielectric constant that also depends on the wavevector. 
 Neglecting the spatial  dispersion  of the dielectric constant  $\kappa^{ex}(\omega, k)$ connected with exciton center of mass motion, one can write the frequency dependent dielectric constant as
\be
\kappa^{ex}(\omega,k=0)= \kappa_{\infty}\bigg(1-\frac{\hbar \omega_{LT}}{\hbar \omega-\hbar \omega_{T}}\bigg)~,
\label{eq:2}
\ee
where $\kappa_{\infty}$ is the high frequency dielectric constant.

After substituting Eq. \eqref{eq:2} in Eq. \eqref{eq:1}, we arrive at  the  expression  that describes the   dispersion of the surface exciton polariton in the semiconductors with very large exciton effective mass:
\be q^{2} =
\frac{(\hbar \omega_T+\epsilon)^{2}}{c^{2}\hbar^2}\bigg(\frac{\kappa_{\infty}(\epsilon-\hbar \omega_{LT})}{\epsilon+\kappa_{\infty}(\epsilon-\hbar \omega_{LT})}\bigg)~,
\label{eq:3}
\ee
where $\hbar$ is the reduced Planck's constant, and $\epsilon$ is the energy photon calculated from the energy of the transverse exciton: $\epsilon= \hbar\omega-\hbar\omega_T$.

Next, we consider how the spatial dispersion of the dielectric constant connected with exciton kinetic energy, $\hbar^2k^2/2M$, which was  neglected in Eq. \eqref{eq:3}, affects the  dispersion of SEP. We were able to find only one paper \cite{LagosPRB1978}  where this problem was considered. The solution was obtained  numerically for SEP on the surface of ZnO. In our work, solving Maxwell's equations\cite{Jackson} and Thomas Hopfield's equations   \cite{HTPR63}, simultaneously, allows us to find an analytical expression for the SEP, which is expressed through a fundamental parameter of exciton/light coupling. We will show that our analytically calculated dispersion is in excellent agreement with the numerical calculations of the ZnO system found in Ref. 4, and that accounting for the exciton dispersion leads to a small, but
quantifiable, modification of the dispersion described by Eq \eqref{eq:3} and small broadening of the SEP line. We will apply our technique to describe the SEP dispersion for several technologically relevant materials and disucss methods of SEP optical excitation and potential use of SEPs in interferometric systems.

{\it Bulk exciton polariton.} To find the SEP dispersion, we first describe exciton polaritons in bulk semiconductors. We consider a semi-infinite semiconductor adjacent to a semi-infinite vacuum, with the $x$-axis perpendicular to the interface and the semiconductor (vacuum) occupying the half space $x > 0 $ ($x < 0$).  
The system of the  equations  here includes the Maxwell equations for electric field $\bm E$ \cite{Jackson}
\bea
{\Delta\bm E} - \nabla(\bm \nabla\cdot \bm E) &=&-{\omega^2\over c^2} \bm D=-{\omega^2\over c^2} ( \kappa_\infty\bm E+4\pi \bm P)\nonumber\\
(\bm \nabla \cdot \bm D)=\bm \nabla \cdot (\kappa_\infty \bm E+4\pi  \bm P)&=&\kappa_0(\bm \nabla \cdot \bm E) + 4\pi(\bm\nabla \cdot \bm P)= 0
\label{eq:ME}
\eea
where $\bm D=\kappa_\infty\bm E+4\pi \bm P$ is the induction vector, and $\bm P$ is the exciton polarization. The  following  equation  was developed by Hopfield and Thomas and describes the motion of the exciton polarization, $\bm P$,  \cite{HTPR63}
\be
-\frac{\hbar^{2}}{2M}\Delta \bm P - \epsilon \bm P = \alpha \bm E
\label{eq:TH} 
\ee
where  $\alpha= \hbar \omega_{LT}\kappa_{\infty}/4\pi$.

Knowing that there  are two types of excitons, longitudinal and transverse, we are looking for the solution where polarization $\bm P$ has following two component form:
\be
\bm P = (P_{x}\bm x_{0} + P_{z}\bm z_{0})e^{iqz+ipx} =\bm P_{0}e^{iqz+ipx} , 
\label{eq:P} 
\ee
where $q$  and $p$ are  the vectors for the wave propagated along  and perpendicular to the interface. Substituting Eq.\eqref{eq:P} into Eq.\eqref{eq:TH}, we obtain the electric field
\bea
\bm E = \bm P_{0}te^{iqz+ip x}
\label{eq:EF}
\eea
where 
\be
t=\frac{1}{\alpha}\bigg(\frac{\hbar^{2}}{2M}(q^{2}+p^2)-\epsilon\bigg) .
\label{eq:9}
\ee

Substituting  the electric field $\bm E$  described by Eq. \eqref{eq:EF} into Maxwell  equations, \eqref{eq:ME} we obtain two equations for $\bm x_{0}$ and $\bm z_{0}$, the projections  of polarization $\bm P_{0}$:
\bea
-t(q^2+p^2)P_x+t(p^2P_x -qp P_z)&=&-{\omega^2\over c^2}(\kappa_\infty tP_x+4\pi P_x)\nonumber\\
-t(q^2+p^2)P_z-t(qp P_x-q^2P_z)&=&-{\omega^2\over c^2}(\kappa_\infty tP_z+4\pi P_z)
\label{eq:9}
\eea
The solution of Eq. \eqref{eq:9} exists if its determinant is equal to zero. This gives us the dispersion equations:
\be
\bigg[(\kappa_{\infty}k_{0}^{2}-q^{2})t+4\pi k_{0}^{2}\bigg]\bigg[(\kappa_\infty k_{0}^{2} -p^{2})t+4\pi k_{0}^{2}\bigg]=q^{2}p^2t^{2} ~,
\label {eq:10}
\ee
where we introduce $k_0=\omega/c$. Equation \eqref{eq:10} can be rewritten using $A=(\kappa_\infty t+4\pi)k_0^2$ and consequently simplified to: $A^2-A(q^2+p^2)t=0$
 that gives us two  solutions. The first one, $A=0$, gives the dispersion of the bulk longitudinal exciton:
 \be
 \epsilon= \hbar \omega_{LT}+ \frac{\hbar^{2}}{2M}Q^2~,
 \ee
 where $Q^2= p^2+q^2$, and the relationship between the projections of the polarization: $P_x=-(p/q)P_z$.

The second one, $A=Q^2 t$,  describes the dispersion for the two branches of the transverse optically active bulk exciton
\be
\frac{\hbar^{2}}{2M}Q^2_{\pm}-\epsilon = \kappa_{\infty}\frac{\hbar^{2}k_{0}^{2}}{4M}-\frac{\epsilon}{2}\pm \sqrt{\left(\frac{\hbar^{2}k_{0}^{2}\kappa_{\infty}}{4M}-\frac{\epsilon}{2}\right)^{2}+\frac{\hbar^{2}k_{0}^{2}}{2M}\kappa_{\infty}\hbar\omega_{LT}} 
\label{eq:12}
\ee
and the relationship between the projections of the polarization: $P_x=(q/p)P_z$.  In Eq \eqref{eq:12}, the momentum $k_0$ is also a function of energy $k_0= (\hbar \omega_T+ \epsilon)/c\hbar$. As an example,  Fig. 1 shows the dispersion of the bulk longitudinal and transverse  excitons calculated for ZnO.

\begin{figure}
\includegraphics[width=16cm]{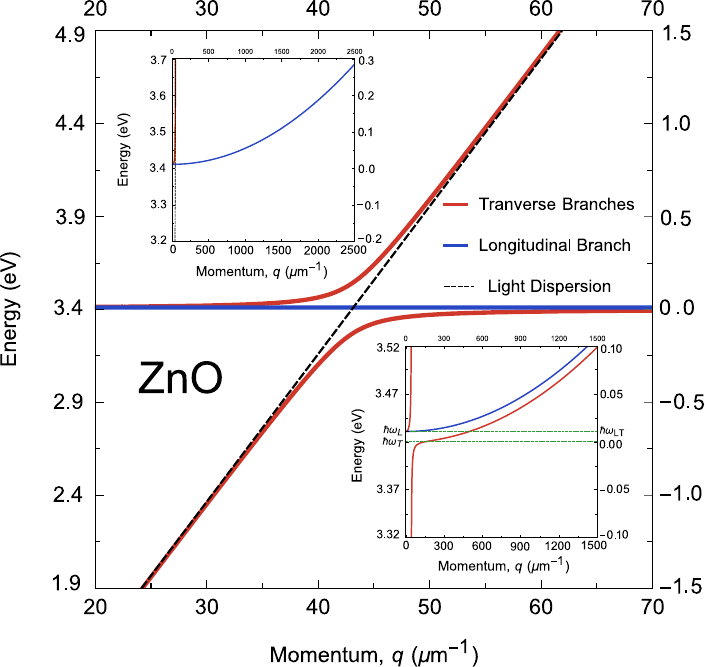}
\caption{Dispersion of the bulk exciton polariton in ZnO semiconductor. The dispersion is  calculated  using parameters from Ref. \cite{LagosPRB1978}:  $\hbar\omega_{L} = 3.4325$\, eV,  $\hbar\omega_{T} = 3.4215$\, eV, $\hbar\omega_{LT} = 0.011$\,eV, $M = 0.87m_0$,   where $m_0$ is the mass of free electron, and $\kappa_{\infty } = 6.16$. The upper inset  shows the dispersion  of  the  bulk exciton polariton at extended momentum range.  The lower inset,  that  magnifies the energy scale  shows the exciton polariton fine structure.}
\label{fig:1}
\end{figure}


{\it Surface exciton polariton. }
Having bulk solutions of Maxwell's equations and exciton polarization in  the semiconductor half-space with $x>0$  and assuming the trivial solution of the Maxwell equation in the vacuum half-space, $x<0$,  we can find the SEP state by satisfying  the Maxwell boundary condition and polarization boundary conditions at the surface $x=0$.  The SEP propagation along the surface is described by the factor $e^{-iqz}$, and it decays exponentially into the semiconductor semi-space $x>0$. This means that we need to replace momentum $p$ by the complex-valued $i\gamma$ in the wave function of the bulk exciton polariton. This also leads  to the  replacement of  $Q^2= p^2+q^2$ by $Q^2= q^2-\gamma_0^2$ in the dispersion of the  longitudinal exciton and  $Q^2_{\pm}=q^2-\gamma_\pm^2$ in the  dispersion of the  transverse  excitons. Any exciton polarization  in the semiconductor  can  be  generally written through 3 independent  polarization  components with decay into the semiconductor described by $\gamma_{0}$,  $\gamma_{-}$ and $\gamma_{+}$:
\bea
P_{z} = \bigg[P_{0}e^{-\gamma_{0}x}+ P_{+}e^{-\gamma_{+}x} +  P_{-}e^{-\gamma_{-}x}\bigg]e^{iqz} \nonumber \\
P_{x} = \bigg[\alpha_{0}P_{0}e^{-\gamma_{0}x}+\alpha_{-} P_{-}e^{-\gamma_{-}x}+\alpha_{+}P_{+}e^{-\gamma_{+}x}\bigg]e^{iqz},
\label{eq:13}
\eea
where in connections between $P_x$ and $P_z$ we replace $p$ by corresponding $i\gamma$. That leads to:
\be \alpha_0=\frac{i\gamma_0}{q} ~,~ \alpha_-= \frac{iq}{\gamma_-}~,~\alpha_+=\frac{iq}{\gamma_+}~.
\label{eq:27}
\ee
\begin{figure}
\includegraphics[width=14cm]{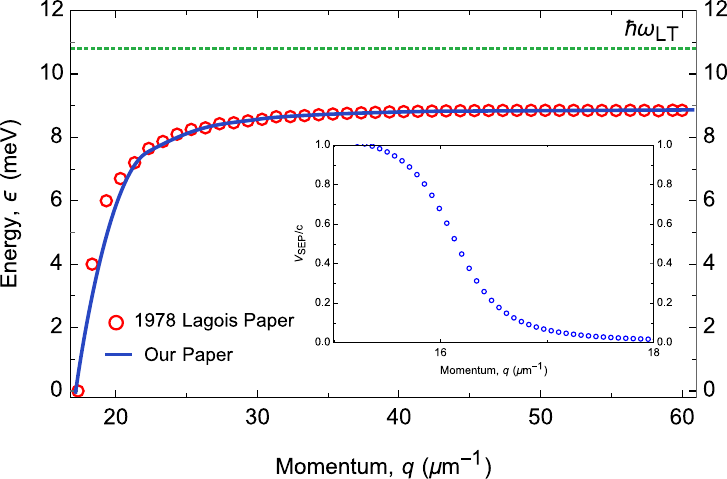}
\caption{Dispersion of the surface  exciton polariton at the  ZnO/vacuum interface.   Our approach was used to calculate the  dispersion (blue line)  using parameters from Ref. 4: $\hbar\omega_{L} = 3.4325$\,eV, $\hbar\omega_{T} = 3.4215$\,eV, $\hbar\omega_{LT} = 11$\,meV, $M = 0.87m_0$,    and $\kappa_{\infty } = 6.16$.  The horizontal  dotted line shows the longitudinal  transverse splitting $\hbar\omega_{LT}$ in ZnO. The red empty circles are the numerical results extracted from  Logos paper.\cite{LagosPRB1978}  Inset shows  the dependence  of  SEP propagation velocity $V_{SEP}/c$  as a function of $q$. Energy scale is relative to the transverse exciton}
\label{fig:2}
\end{figure}
In our paper we use the Pekar additional boundary conditions (ABCs), which states that exciton polarization vanishes at the semiconductor surface,
\be
 P_0+ P_++P_-= 0~,~  \alpha_0  P_0 + \alpha_+  P_+ + \alpha_-P_- = 0~,
\label {eq:33}
 \ee
The ABCs allow to rewrite $P_{\pm} $ via $P_0$.
\be 
 P_+ = \frac{\alpha_- - \alpha_0}{\alpha_+ - \alpha_-} P_0~, ~ P_- = \frac{\alpha_0 - \alpha_+}{\alpha_+- \alpha_-}P_0
\label{eq:34}
\ee
\begin{figure}
\includegraphics[width=13cm]{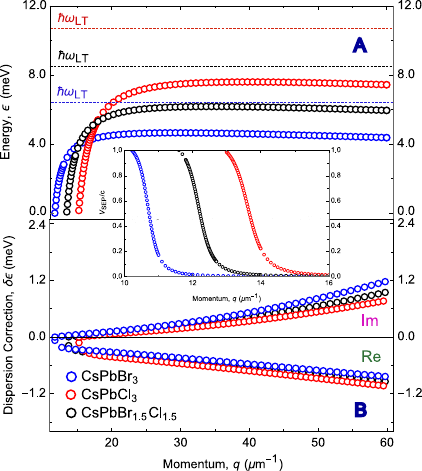}
\caption{Dispersion of SEP in CsPbBr$_3$ and CsPbCl$_3$ and CsPbBr$_{1.5}$Cl$_{1.5}$ (panel A) Calculations for CsPbBr$_3$ were conducted using following parameters: $\hbar\omega_{L} = 2.3153$\,eV , $\hbar\omega_{T} = 2.3090$\,eV \cite{Transverse},  and $\hbar\omega_{LT} = 6.3$\,meV \cite{Measurement};  $M = 0.214m_{0}$ \cite{Mass},  and $\kappa_{\infty}=4.96$.\cite{Measurement}
Calculations for CsPbCl$_3$ were done using parameters  $\hbar\omega_{L} = 3.0027$\,eV, $\hbar\omega_{T} = 2.9920$\,eV,\cite{Transverse}  and $\hbar\omega_{LT} = 10.7$\,meV,\cite{Measurement};  $M = 0.202m_{0} $ \cite{Mass}, and  $\kappa_{\infty} = 4.07$.\cite{Measurement}. Parameters for CsPbBr$_{1.5}$Cl$_{1.5}$ were found by averaging the parameters in CsPbBr$_3$ and CsPbCl$_3$.  The horizontal  dotted line show the longitudinal  transverse splittings $\hbar\omega_{LT}$, in corresponding perovskite compounds. Panel  B shows the corrections connected with final effective mass.  Inset  shows the SEP propagation  velocity $V_{SEP}/c$  in considered perovskites as a function  of quasi momentum $q$. Energy scale is relative to the transverse exciton}
\label{fig:3}
\end{figure}
 Substituting  Eq. \eqref{eq:13} into  Eq. \eqref{eq:EF}, we obtain expressions for two components of the electric field:
\bea
E_{z} = \frac{1}{\alpha}\bigg[-\hbar\omega_{LT}P_{0}e^{-\gamma_{0}x}+(\delta+\sqrt{\delta^{2}+\tau})P_{+}e^{-\gamma_{+}x}+(\delta-\sqrt{\delta^{2}+\tau})P_{-}e^{-\gamma_{-}x}\bigg]e^{iqz}\nonumber\\
E_{x} =\frac{1}{\alpha}\bigg[-\hbar\omega_{LT}\alpha_{0}P_{0}e^{-\gamma_{0}x} + (\delta+\sqrt{\delta^{2}+\tau})\alpha_{+}P_{+}e^{-\gamma_{+}x}+(\delta-\sqrt{\delta^{2}+\tau})\alpha_{-} P_{-}e^{-\gamma_{-}x}\bigg]e^{iqz} ~,\nonumber\\
\label{eq:EF2}
\eea
where  $\delta= \kappa_{\infty}(\hbar^{2}k_{0}^{2}/4M)$ and $\tau=\kappa_{\infty}\hbar\omega_{LT}(\hbar^{2}k_{0}^{2}/2M)$.

The electric fields  $E_x$ and $E_z$  should satisfy the standard  boundary conditions of continuity of the tangential  component $E_z$ and the  induction vector's normal component $D_x=\kappa_\infty E_x$.
On the other side of the half space, $x<0$,  the electric field of the propagating wave  decaying in vacuum can be written as:
\be
\bm E = \bm R e^{iqz+\gamma x}=(R_{x}\bm x_{0} + R_{z}\bm z_{0})e^{iqz+\gamma x}~,
\ee
where  $q^{2}-\gamma^{2}=k_{0}^{2}$. Maxwell's equation in vacuum  $\nabla \cdot \bm E=0$  gives: $iqR_{z} + \gamma R_{x}=0$.  Consequently  the ratio:
$R_x/ R_z=-iq/ \gamma= -iq/ \sqrt{q^2-k_0^2}$.  This allows one to write the boundary conditions on electric field components:
\be
{\kappa_\infty E_x\over E_z}={R_x\over R_z}={-iq\over  \sqrt{q^2-k_0^2}}
\label{eq:19}
\ee
Substituting $E_x$ and $E_z$  from Eq. \eqref{eq:EF2} into Eq. \eqref{eq:19} we arrive  at an equation that describes the dispersion of SEP:
\bea
&&\hbar\omega_{LT}\left(\gamma_0+{q^2\over\kappa_\infty \sqrt{q^2-k_0^2}}\right)(\gamma_--\gamma_+)  -(\delta+\sqrt{\delta^{2}+\tau})(q^2-\gamma_0\gamma_-)\left(1+{\gamma_+\over\kappa_\infty\sqrt{q^2-k_0^2}}\right)\nonumber\\
&-&(\delta-\sqrt{\delta^{2}+\tau})(\gamma_0\gamma_+- q^2)\left(1+{\gamma_-\over\kappa_\infty\sqrt{q^2-k_0^2}}\right)=0
\label{eq:SEP}
\eea
This equation generally does not have an analytical solution.  

One notices, however,  that  we are interested in  exciton dispersion   for small  $q$ which satisfies the condition $\hbar^2q^2/2M\ll \hbar\omega_{LT}$ and in wide band-gap semiconductors, we investigate $\hbar^2k_0^2/2M\ll \hbar\omega_{LT}$. In this case, one can make an  expansion  of Eq. \eqref{eq:SEP} in the power series of large mass $M$. This leads to:
\be
\gamma_0\gamma_+A(\epsilon, q)+ B(\epsilon, q)=0~,
\label{eq:Exp}
\ee
where $\gamma_0 \sim \sqrt{M}$,  $\gamma_+ \sim \sqrt{M}$ and  the leading term in $ B(\epsilon, q)\sim \sqrt{M}$ as it is shown in SI.
In the first approximation, the leading term in Eq. \eqref{eq:Exp}  must vanish: $A(\epsilon, q)=0$.  Using the same approximation  that $\hbar^2q^2/2M\ll \hbar\omega_{LT}$ and $\hbar^2k_0^2/2M\ll \hbar\omega_{LT}$  we show (see SI) that:
\be
A(\epsilon, q)=\frac{-\sqrt{q^2-\kappa(\epsilon)k^2_0}\cdot\epsilon}{\kappa_{\infty}\sqrt{q^{2}-k_{0}^{2}}}-\hbar\omega_{LT}+\epsilon=0~,
\label {eq:A}
\ee
where  we introduce $\kappa(\epsilon)=\kappa_\infty(\epsilon-\hbar\omega_{LT})/\epsilon$. Rewriting Eq. \eqref{eq:A} as $\sqrt{q^2-\kappa(\epsilon)k^2_0}=-\kappa(\epsilon)\sqrt{q^{2}-k_{0}^{2}}$ and
squaring  left and right sides, we arrive at  Eq: \eqref{eq:3} that describes  SEP surface states in the absence of exciton dispersion:
\be
q^{2} = k_{0}^{2}\frac{\kappa(\epsilon)}{\kappa(\epsilon)+1}~.
\label{eq:23}
\ee

Now let us find the corrections  to  $\epsilon$  connected  with the finite exciton mass using  Eq. \eqref{eq:Exp}. Equation \eqref{eq:Exp} can be rewritten as $A(\epsilon, q)+ \theta(\epsilon, q)=0$,  where  $\theta(\epsilon, q)=B(\epsilon, q)/\gamma_0\gamma_+$.  Using a standard Taylor expansion, we can write: $\partial A(\epsilon,q)/\partial \epsilon|_{\epsilon=\epsilon_0}\delta\epsilon=-\theta(\epsilon_0,q)$, where $\epsilon_0$ is the solution of equation $A(\epsilon_0, q)=0$.  That gives the correction to the SEP dispersion $\delta\epsilon$:
\be
\delta\epsilon= -\theta(\epsilon_0,q)/\partial A(\epsilon,q)/ \partial \epsilon|_{\epsilon=\epsilon_0}~.
\label{eq:24}
\ee
An analytical expression for $\theta(\epsilon,q)$ and $\partial A(\epsilon,q)/ \partial \epsilon$  based  on  material parameters  of the semiconductor allows  us to find  SEP dispersion for any semiconductors with relatively large longitudinal-transverse splitting.  In Fig. 2 we show the SEP dispersion at the surface of ZnO, which is in excellent agreement with numerical calculations of Ref. 4.
The dispersion curve shown in Fig. \ref{fig:2} can be easily  understood   if your rewrite Eq. \eqref{eq:23} in the following form:
\be
\epsilon = {\hbar\omega_{LT}(q^2/k_0^2-1)\over (1+1/\kappa_\infty)q^2/k_0^2-1}
\label{eq:79}
\ee
One can see from Eq. \eqref{eq:79} that this dispersion starts from  $\epsilon=0$ when $q=k_0$  goes to asymptotic value $\epsilon=\hbar\omega_{LT}\kappa_\infty/(1+\kappa_\infty)$ at $q\rightarrow \infty$  if we neglect $\delta \epsilon$ corrections described in Eq. \eqref{eq:24}. 

In Fig. \ref{fig:3} panel A, we have calculated the complete SEP dispersion in CsPbBr$_3$ and  CsPbCl$_3$ perovskites and their 50\% alloy CsPbCl$_{1.5}$Cl$_{1.5}$. One can see that, indeed, the SEP  dispersion  starts from $k_0$ for the corresponding perovskite. Due to the increase of the energy gap, $k_0$ in CsPbCl$_3$  is larger than $k_0$ in  CsPbCl$_{1.5}$Cl$_{1.5}$ and  CsPbBr$_3$. One can see also the asymptotic behavior of   $\epsilon$ which goes to $ \hbar\omega_{LT}\kappa_\infty/(1+\kappa_\infty)$  at large $q$. The values of the longitudinal transverse splitting $\hbar\omega_{LT}$, for corresponding compounds are show by the dotted horizontal lines. 

Equation Eq.\eqref{eq:79} also allows for convenient calculations of SEP propagation  speed:
\be
V_{SEP}={1\over \hbar} {\partial \epsilon\over \partial q}+{1\over \hbar} {\partial \delta\epsilon\over \partial q}~,
\ee
where the first term  has the analytical expression for corrections $\delta\epsilon\sim q^2$. One can see that $V_{SEP} =c$ at $q=k_0$, and at large $q$, the first term disappears, but the second term  is proportional to $q$.  In Fig. 2 and 3, we show the dependence of  $V_{SEP}$ as a function of $q$.  One can see from the Insets in these figures  that  the dozen fold decrease of the wave  propagation speed could be reached in SEP relative to the light  with the same frequency if we could excite SEPs with rather large momentum $q$. 
\begin{figure}
\includegraphics[scale = 0.63]{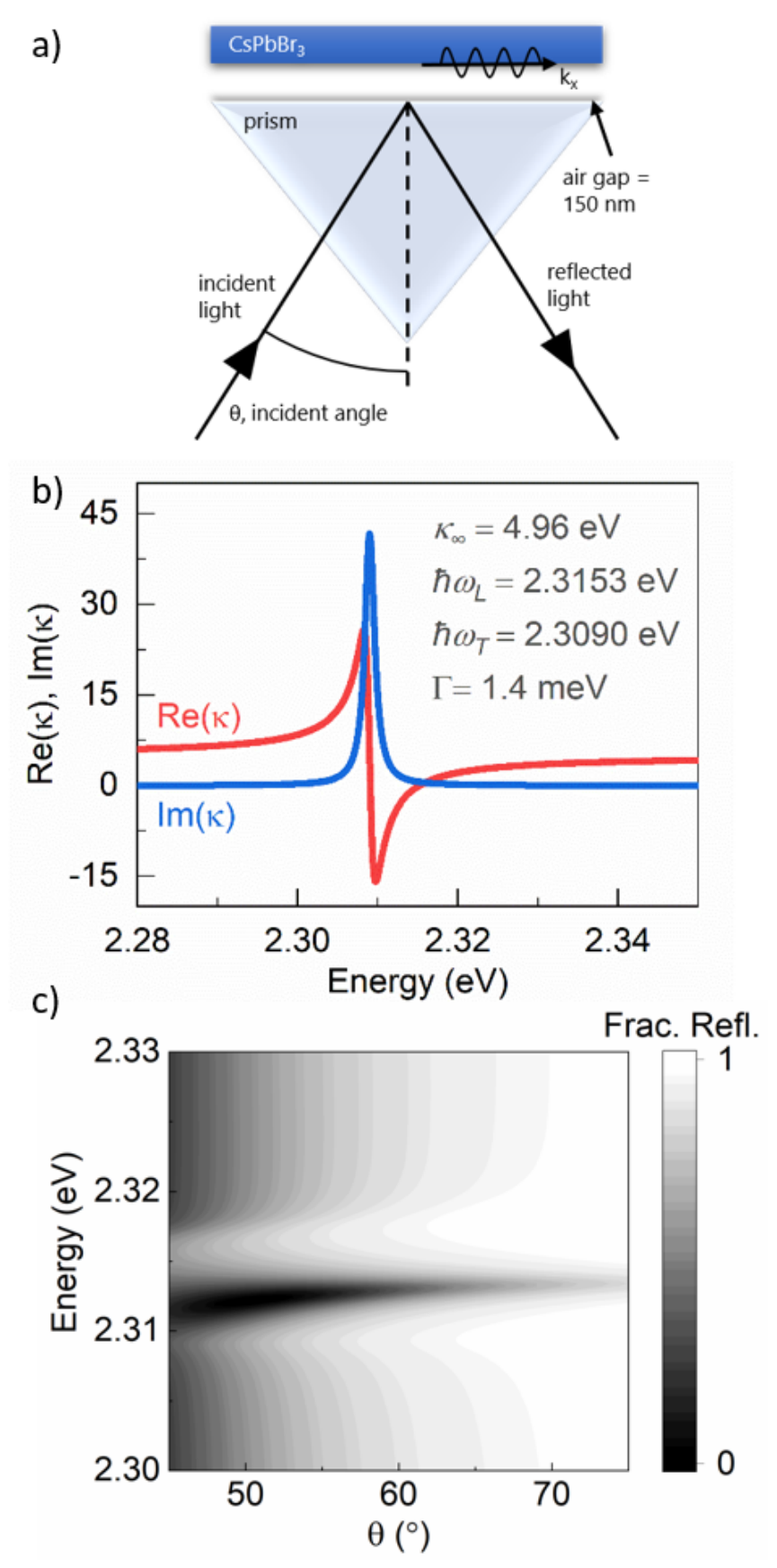}
\caption{a) Schematic of Otto-type prism coupling for exciting and probing surface polaritons. b)
Calculation of the dielectric constant for CsPbBr3 using parameters from Figure 3 and $\Gamma$ = 1.4 meV. c)
Example of the polariton dispersion calculated for surface exciton polaritons supported on CsPbBr3 with
the dielectric constant shown in b). The calculation assumes the CsPbBr3 is positioned 150 nm away from
a prism with a refractive index of 1.475. This contour plot shows a reflection dip (black line), corresponding to polariton formation, that disperses with incident angle.} 
\label{fig:4}
\end{figure}

SEPs can be launched, and their dispersion monitored, using prism-coupling  in the Kretschmann-Raether, Otto configurations, or grating coupling. These methods overcome the inherent momentum mismatch between free-space light
and the polariton mode. For the Otto-type prism-based methods, (see Fig. \ref{fig:4} a), one positions the excitonic material very close to the surface of a prism. Broadband or monochromatic light is then incident through  the prism and onto the excitonic material at some angle and spectral reflection is monitored. Successful coupling of light into the propagating SEP mode is indicated by a dip in the reflected spectrum. Such an approach has been used extensively in the field of plasmon polaritons and to characterize SEPs in inorganic materials \cite{LFExp,Epstein20} and organic dyes.\cite{dies,dies1,dies2, last2}. 

For excitonic material, the dielectric constant $\kappa(\omega)$ goes below -1 only at low
temperatures. For instance, it has been shown that SEPs can be supported in 2D layers of  TMD when encapsulated with hexagonal-boron-nitride and cooled to cryogenic temperatures \cite{Epstein20}. In addition, we expect that SEPs can be supported in CsPbBr$_3$ single crystals at low temperatures. Neglecting spatial dispersion, the dielectric constant of exciton polaritons can be estimated as
\be
\kappa(\omega, k) = \kappa_{\infty} + \frac{\kappa_{\infty}(\omega_{L}^{2}-\omega_{T}^{2})}{\omega_{T}^{2}-\omega^{2}-i\omega\Gamma/\hbar}
\ee
where $\Gamma$ is a damping constant.\cite{last2} It has been shown that $\Gamma$ $\sim$ 1.4 meV in single crystal CsPbBr$_3$ at temperatures below $\sim$ 30K.\cite{last} In Fig. 4b, we calculate the CsPbBr$_3$ dielectric constant using the parameters from Fig. 3 and $\Gamma$ = 1.4 meV, and show the dielectric constant does become negative between the transverse and longitudinal excitons. Using this dielectric function, we calculate  the  angular dependent reflection  for the experimental configuration shown in Fig. 4a. The calculated dispersion (Fig. 4c) shows the reflection deep associated with SEP formation and demonstrates the potential of CsPbBr$_3$ perovskites to support SEPs.

In summary, we have developed a theory describing generation of light-matter quasi-particle SEPs, which accounts for the spatial dispersion of the dielectric constant associated with exciton motion.  Our analytical theory produces an excellent agreement with the only numerical  modeling of this problem  conducted  for SEPs at a  ZnO/vacuum interface. Our calculations do reveal a rather small broadening of these photon-like particles  due to spatial dispersion of the dielectric response that allows for coherence transfer  over long distances.  In combination with very small propagation velocities of SEPs, one can increase the  sensitivity of some coherent devices, such as Sagnac inteferometer applicable to navigation technologies.

{\it Acknowledgments,}  J. H. acknowledges the  financial support  of  SEAP for  his  2024 summer   fellowship  at  US Naval Research Laboratory.  J. O. and Al. L. E  acknowledge  partial support  of  Office of the Undersecretary of Defense (Research \& Engineering) ARAP program. D. R.,  B.S., and    Al. L. E. acknowledge support of the US Office of Naval Research (ONR) through the Naval Research Laboratory’s Basic Research Program.

\end{document}